\documentclass[aps,prl,twocolumn,superscriptaddress,longbibliography]{revtex4-1}%
\usepackage{graphicx} 
\usepackage{float} 
\usepackage{dcolumn}
\usepackage{bm,color}
\usepackage{amsmath,amssymb,dsfont,amstext,amsfonts}
\usepackage{extarrows}
\usepackage[colorlinks=true,linkcolor=blue,urlcolor=blue,citecolor=blue]{hyperref}
\usepackage{xcolor}
\usepackage{wasysym}
\usepackage{mathtools}
\usepackage{bbold}

\usepackage[normalem]{ulem} 
\definecolor{AB}{rgb}{255,0,255}

\begin{document}
\title{Comment on ``Floquet non-Abelian topological insulator and multifold bulk-edge correspondence"}
\author{Robert-Jan Slager}
\affiliation{TCM Group, Cavendish Laboratory, University of Cambridge, JJ Thomson Avenue, Cambridge CB3 0HE, United Kingdom\looseness=-1}
\author{Adrien Bouhon}
\affiliation{TCM Group, Cavendish Laboratory, University of Cambridge, JJ Thomson Avenue, Cambridge CB3 0HE, United Kingdom\looseness=-1}
\author{F. Nur \"{U}nal}
\affiliation{TCM Group, Cavendish Laboratory, University of Cambridge, JJ Thomson Avenue, Cambridge CB3 0HE, United Kingdom\looseness=-1}
\begin{abstract}
We comment on the recent paper ``Floquet non-Abelian topological insulator and multifold bulk-edge correspondence" by 
Tianyu Li and Haiping Hu, Nat. Comm. {\bf 14}, 6418 (2023). Apart from the fact that the authors unjustly imply to study multi-gap topology in Floquet systems for the first time, only known homotopic relations are presented. While such insights are used to present interesting Floquet phenomena and phases, which is an attractive result in itself, they cannot be used to deduce the total bulk characterization in the dynamical context without further proof. In fact, the authors essentially rephrase a Zak phase description. These results should in particular be contrasted to earlier results, arXiv:2208.12824, in which static-compatible Zak  phases {\it and} dynamical Dirac strings were shown to be able to {\it distinguish} rather similar non-Abelian Floquet phases in $2+1$ dimensional systems. As a result, the claim of a sharp multifold bulk-edge correspondence cannot be concluded from the given arguments. 
\end{abstract}
\maketitle
{\it Introduction---} In a recent paper Li and Hu (L-H) investigate multi-gap topology in Floquet systems~\cite{lihu2023}. Multi-gap topologies concern an upcoming field in which more general band partitions can lead to isolated band subspaces,\;i.e.\;separated by energy gaps from above and below, having invariants beyond the  conventional occupied-unoccupied band partitioning~\cite{bouhon2020geometric, bouhon2019nonabelian, BJY_nielsen}. A prime example is Euler class arising in systems that allow a choice of basis in which the Bloch Hamiltonian is a real matrix for all momenta. Nonzero Euler class can be hosted by any two-band subspace and is characterized by the presence of a number of band degeneracies that cannot be removed as long as the two energy gaps (below and above) are preserved. Crucially, these band nodes can be alternatively described by non-Abelian frame charges associated with frames of Bloch eigenvectors over a loop encircling these band degeneracies \cite{Wu1273,bouhon2019nonabelian}.
Furthermore, by braiding band degeneracies of {\it adjacent} energy gaps around each other in momentum space, non-Abelian frame charges can be altered to induce a two-band subspace with similarly-valued degenracies, whose obstruction to annihilate is quantified by evaluating the Euler class over a Brillouin zone patch that contains the charges of this two-band subspace~\cite{bouhon2019nonabelian}. 

Although it was already reported that such multi-gap topological evaluations can result in new observables and phases in out-of-equilibrium contexts~\cite{Unal_quenched_Euler} and Floquet systems~\cite{slager2022floquet} in particular, the study of new interplays constitutes a promising direction. In this regard, translating  static topological characterizations to Floquet systems {\it in a gap specific manner} entails an effective strategy to characterize new dynamical topological phases that may not have any static counterpart. Indeed, the extra gap stemming from the time periodicity at the edge of the Floquet Brillouin zone (FBZ) can be identified by connecting to the high-frequency regime~\cite{Unal19_PRL} where the Floquet replicas are well separated. 
This analysis in fact has been used for gap specific measurements of the winding numbers in  anomalous Floquet topological phases~\cite{Unal19_PRL,Wintersperger20_NatPhys}. In the multi-gap topological context, this argument may be continued directly as it is known that the quaternion group generalizes to the Salingaros’ vee group~\cite{Wu1273,bouhon2019nonabelian} for more bands. Similarly, partial-frame charges~\cite{Peng2021} may be defined, further showing that one can effectively ``embed'' the static insights in the replica Floquet band structure. As in Ref.~\cite{slager2022floquet}, it is this strategy Li and Hu use. While we cannot agree with some of their essential conclusions, we appreciate the new phases, in particular the Floquet non-Abelian topological insulator having edge states in all gaps, that are obtained in this manner.

{\it Back ground and intial comments---} In contrast to Ref.~\cite{slager2022floquet}, L-H restrict to $1+1$D systems without braiding and, hence, cannot identify Euler class. This naturally puts more emphasis on the definition of $1$D charges as elements of the quaternion group due to natural identification of the eigenstates (or in phase bands in the Floquet context) with an orthonormal dreibein. This situation is quite analogous to the difference between the recent meta-material realizations of multi-gap non-Abelian phases in Ref.~\cite{Jiang2021}
and Ref.~\cite{Guo1Dexp}, where the former considered 2D braiding resulting in a finite Euler class and the latter realized the quaternion  charges in 1D. While the periodicity of $1$D Brillouin zones come with some inherent peculiarities that should be addressed to formally define parallel transport and thus definition of quaternion $1$D charges, this is a subject that deserves  precise  mathematical treatment which we will report on in detail in a separate paper soon.

Although these subtleties should, thus, already be taken into account to define charges in the static context, of direct present concern is that L-H claim to derive a dynamical bulk-boundary correspondence (BBC) and thus a full quantification of the bulk invariants or 1+1D charges.
The main problem with this proposed BBC however is (as we will detail below) that distinct phases are labelled by the same invariant. Rather than extending the invariant, this problem is reformulated by L-H as a ``multifold BBC". To demonstrate the shortcomings of this claim, a brief recap of literature would be useful. 

L-H use the concept of phase bands~\cite{Nathan15_NJP} and singularities therein that have been successfully employed to classify conventional single-gap topologies~\cite{Unal19_PRL}, including the anomalous Floquet topological insulator (AFTI)~\cite{Rudner13_PRX}. The latter considers a two-band Floquet spectrum that due to the double band inversion [one through the static energy gap and one through the edge of the FBZ] has zero Chern number, but nonetheless exhibits an edge state in each gap, reflecting the double band inversion. 
As L-H focus on $1+1$D systems, the phase bands are indexed by one momentum and a time parameter and, hence, nodal points carry charges indexed by the first homotopy group. Here, L-H can thus repeat the calculations of Refs.~\cite{bouhon2020geometric,Wu1273, bouhon2019nonabelian} and evaluate $\pi_1(M_3)$, with $M_3=O(3)/O(1)^3$, to get the  quaternion charges.
L-H then use these phase band charges to define a total bulk charge
\begin{equation}\label{eq:winding}
q=\prod_n \tilde{q}_n,    
\end{equation}
where $\tilde{q}_n$ are the phase band charges in the $n^{\text{th}}$ gap.

We first like to point out that the above description is very much a Zak phase analysis [albeit gap dependent] in disguise. Indeed, all the non-Abelian charges (except for the quaternion element `$-1$') directly relate to Zak phases, as indicators of the homotopy classes of the fundamental group (first homotopy group). The first homotopy charges can physically be seen as $\pi$-vortices in each direction of the frame spanned by the eigenstates or phase bands. In Ref.~\cite{slager2022floquet}, such Zak phase descriptions for multi-gap (non-Abelian) Floquet phases were detailed for the first time and it was already observed that there are distinct dynamical phases that, while hosting the same set of Zak phase per band, exhibit different edge spectra [see Methods in Ref.~\cite{slager2022floquet}, pages 7 and 8 in particular]. In that work, a 2D three-band example was considered in which one starts with Zak phase projections $(\gamma_{1}, \gamma_{2}, \gamma_{3})=(0,\pi,\pi)$ for bands 1, 2 and 3, along one of the reciprocal lattice vectors. Assuming this entials an atomic limit with no edge states, two processes involving either a band inversion in the anomalous gap or, alternatively, two consecutive band inversions in the two regular gaps gives the same final configuration $(\gamma_{1}, \gamma_{2}, \gamma_{3})=(\pi,\pi,0)$. The first process however gives an edge state spectrum in the anomalous gap, whereas the second a double edge state in the other two gaps. This represents the $ij=k$ multiplication. Similarly, it is straightforward to see that e.g.~the first column in Fig.3 of L-H connecting the Floquet non-Abelian topological insulator and trivial phase with no edge states follows an analogous procedure [i.e.~doing a band inversion in all gaps giving the same configuration of Zak phases as each band undergoes two band inversions accounting for a total change of Zak phase $(\pi+\pi) = 0 \mod 2\pi$]. We, however, emphasise the need to perform band inversions to connect these phases.

\textbf{\emph{Statement of critique---}} As such, we can see the concerns with L-H's claims. First of all, they assume descriptions that are only proven when the Zak phases are a faithful description of the system, while already in the static case unexplained edge states have been observed~\cite{Jiang2021, Peng2021}. Secondly, and more profoundly, there is no direct evidence yet that the different phases that amount to the same $q$ do not have a different characterization that separates them and explains the different edge spectra. In this sense, it is peculiar that on one hand, phases are considered different [such as the trivial and Floquet non-Abelian topological insulator that are both characterized by $q=1$], but because they have the same $q$ it is claimed that this implies a ``multifold bulk-boundary correspondence''. One might also interpret the difference in the edge spectra as a clear indication that the `$q$' charges used in Ref.~\cite{lihu2023} are not sufficient to classify all dynamical Floquet phases. Indeed, to go from one phase that is compatible with a static limit, to its dynamic counterpart, i.e.~exhibiting the same `$q$' invariant (and thus the same Zak phases), the phase bands must undergo inversions, implying that they cannot belong to the same homotopy class. In other words, one cannot transform one phase into the other adiabatically, seemingly contradicting the claim of Ref.~\cite{lihu2023}. In contrast, even if the charges were exhaustive, this would mean that the phases in the left and right column of Fig.3 can be deformed into each other with symmetry preserving terms. This would imply that edge states can be removed and makes the Floquet non-Abelian topological insulator non-topological. At the very least, this point must be addressed to claim a full BBC.  In fact, a proof that the defined charges exhaustively classify the bulk topology would have been desired.

{\it Further illustration---} To make the above criticism more concrete, we note that the charges are only non-Abelian with respect to the other gaps/states of the frame. Hence, when considering per gap, we can draw analogies to the established single-gap Abelian cases. For example, the AFTI phase not amended by any symmetries other than translations has been shown to be captured by winding numbers that involve both momentum and time~\cite{Kitagawa10_PRB,Rudner13_PRX}. Due to the double band inversion through the regular energy gap [that can be adiabatically connected to a static energy gap] and through the anomalous energy gap, the Chern numbers are successively converted from zero to finite values, and then back to zero. This leaves a chiral edge state running across each gap while the total Chern number for both the trivial phase and AFTI are zero. Nonetheless, the two can be distinguished by a {\it complete} bulk-boundary correspondence~\cite{Rudner13_PRX}, rather than interpreted as different edge state spectra of this same invariant implying a multifold BBC. 
Moreover, in the full BBC of the AFTI, the winding numbers faithfully capture the edge spectra by simultaneously extending the 2D Chern invariant to {\it truly} 2+1D topological invariants and considering gap specific data corresponding to different branch cuts. Indeed, there are different routes, also using Hopf maps~\cite{Unal2019}, to prove that the winding numbers address both the regular and anomalous gaps.

At the moment, it is not clear whether there is a formulation involving momentum and time such that both the non-Abelian charges and the gap-dependent band inversions can be accounted for. Nevertheless, as pointed out above, the distinct edge spectra obtained in Ref.~\cite{lihu2023} is an indication of a richer dynamical topology. In this regard, it would have been more prudent to demonstrate the {\it phases} found by using the gap-dependent evaluation, rather than claiming a full profound BBC. Here, we note that a priori it seems that, similar to insightful earlier strategies explored in Ref.~\cite{Roy17_PRB}, the analysis of unitary loops that include contributions that do not connect to identity can shed light on the anomalous non-Abelian topological phases [listed in the right column in Fig.\;3 of Ref.~\cite{lihu2023}]. 

From an edge perspective, we note that the domain wall construction of Ref.~\cite{Guo1Dexp} suffers robustness issues already in the static case. Indeed, a faithful BBC of multi-gap phases is still an active area of research. So far only Zak phases have been successful in describing the edge states in some cases~\cite{jiang_meron}, but a full BBC has yet to be formulated. The domain wall construction only works if it coincides with faithful Zak phases' analysis, which are known not to account for all edge states~\cite{Jiang2021,Peng2021}, or when the different Hamiltonian on either side relate to the same $H(k=0)$. This is detailed with examples in the peer review report of Ref.~\cite{Guo1Dexp}, see pages 39, 40 and 41 in particular. L-H claim it works generally for the dynamical case, but a general proof is not given.

{\it Comparison with arXiv:2208.12824 and comments on presentation---} Finally, we point out that some parts of the presentation of the paper are slightly misleading as already alluded to above. The paper was received on the 14th of April 2023 and by referring to arXiv:2208.12824 (Ref.~\cite{slager2022floquet}), L-H acknowledge awareness of this work that appeared in August 2022. Illustrated by the above examples, some of the questions and strategies of L-H directly coincide with Ref.~\cite{slager2022floquet}. In this regard, statements as 
``{\it Our work, for the first time, presents Floquet topological insulators characterized by non-Abelian charges and opens up exciting possibilities for exploring the rich and uncharted territory of non-equilibrium topological phases}" are suggestive. Indeed, the $2+1$ dimensional phases of Ref.~\cite{slager2022floquet} involve multi-gap dependent non-Abelian charges and furthermore their braiding under periodic driving (c.f.~the main questions addressed by L-H). This suggestive presentation is also reflected in unjustifiable answers to referee questions, where L-H state to ``{\it for the first time study Floquet non-Abelian topological phases emerging from multiple tangled band gaps}'' and the manner in which Ref.~\cite{slager2022floquet} is referred to. In fact, until the outlook/discussion section, no mention of Ref.~\cite{slager2022floquet} is made other than citing in a collection of papers that defined multi-gap topology in the undriven context. In the discussion, it is then mentioned that the $2+1$D classifying space $\mathbb{RP}^2$ is relevant for the Floquet Euler phase. Here we note that somehow the authors have removed ``anomalous'' from the nomenclature. 
L-H might interject here that in the anomalous Dirac string phase and anomalous Euler phase introduced in Ref.~\cite{slager2022floquet}, the spectrum only features an edge state in one gap. Here a concern might be that, while the edge states are hosted in the anomalous gap, one might shift the spectrum and then question the anomalous nature. We would like to point out in this regard that in those models a Kagome geometry was purposefully chosen. 
Indeed, the anomalous Dirac string phase and anomalous Euler phase in fact host a {\it Dirac string} or an Euler class {\it in each gap} of their Floquet spectrum. As such, employing the definition that an anomalous phase is one in which all gaps including the one at the FBZ edge are needed to characterise the topological nature of the phase justifies the term anomalous in a natural manner.

We can make this anomalous nature more precise from another angle. In particular, the anomalous phase presented in Ref.\;\cite{slager2022floquet} with all bands fully gapped (i.e. the flag limit) ---a phase that was named the ``anomalous Dirac string phase''--- exhibits an edge state in the anomalous gap, while the regular gaps have no edge state but still host Dirac strings. By tracking the Dirac string configuration, any procedure to shift the edge state from the anomalous gap to others or to connect to the trivial phase (representable by an atomic limit) can be proven to involve band inversions~\cite{slager2022floquet}. This similarly holds for the Anomalous Euler phase that cannot be compatible with a static counterpart. 
We strongly emphasise that the Dirac strings are the homotopy invariants that keep track of the gap-dependent band inversions in a dynamic Floquet system and are complementary to the Zak phases defined by using band eigenstates. Crucially, this effectively resolves any apparent ``multifold'' BBC that would have been inherited if solely Zak phases were considered.

We have argued above that a full bulk classification for the 1+1D case addressed in \cite{lihu2023} remains yet to be proven and at the very least should include reflection on the above arguments. This puts the multi-fold BBC concept into question when based solely on the present arguments. 

{\it Acknowledgements} R.-J.S. acknowledges funding from a New Investigator Award, EPSRC grant EP/W00187X/1, a EPSRC ERC underwrite grant  EP/X025829/1, and a Royal Society exchange grant IES/R1/221060 as well as Trinity College, Cambridge. A.B.~was funded by a Marie-Curie fellowship, grant no. 101025315. F.N.\"U.~acknowledges funding from the Marie Sk{\l}odowska-Curie programme of the European Commission Grant No 893915, Trinity College Cambridge.

\bibliography{references}

\end{document}